# Optical control of in-plane domain configuration and domain wall motion in ferroelectric and ferroelastic

Vivek Dwij[1], Binoy Krishna De[1], Hemant Singh Kunwar[1], Sumesh Rana[1], Praveen Kumar Velpula[1], D. K. Shukla[1], M. K. Gupta[2,3], R. Mittal[2,3], V. G. Sathe[1*]

[1] UGC-DAE Consortium for scientific research, Indore
[2] Solid State Physics Division, Bhabha Atomic Research Center, Mumbai 400 085, India
[3] Homi Bhabha National Institute, Anushaktinagar, Mumbai 400094, India
*Email: vasant@csr.res.in

**Abstract**

The sensitivity of ferroelectric domain walls to external stimuli makes them functional entities in nanoelectronic devices. Specifically, optically driven domain reconfiguration with in-plane polarization is advantageous and thus highly sought. Here, we show the existence of in-plane polarized sub-domains imitating a single domain state and reversible optical control of its domain wall movement in a single-crystal of ferroelectric $BaTiO_3$. Similar optical control in the domain configuration of non-polar ferroelastic material indicates long-range ferroelectric polarization is not essential for the optical control of domain wall movement. Instead, flexoelectricity is found to be an essential ingredient for the optical control of the domain configuration and hence, ferroelastic materials would be another possible candidate for nanoelectronic device applications.

**Introduction**

Multifunctional ferroelectric devices with high-density data storage and efficient sensing capabilities [1] are demand of modern information technology. A switchable bipolar ground state of ferroelectrics is indispensable for low-energy electronic devices, however, limited to binary switching [2]. The ferroelectric state is generally accompanied by exotic dipolar arrangements and hence contains a complex domain structure [3-4]. Precise control and manipulation of ferroelectric domain walls (DWs) movement offer a new data storage principle where data is stored and processed by the DWs [5,6]. The possibility of continuous tunability of the DWs can extend the concept of binary state to multistate switching [7]. This makes reconfigurable ferroelectric DWs vital for the next generation of nanoelectronics [8,9,10]. The main hurdle in using ferroelectrics is the destabilization of ferroelectric polarization at the low dimensions, suppression of the ferroelectric transition due to the increased depolarization fields, and currently employed charge sensing-based destructive readout schemes [5,11]. Most of the previous studies focused on controlling domain configurations with up/down, intercorrelated polarization [12,13] while domain configurations with purely in-plane polarized domains and DWs remained elusive [14] because of the inherent difficulties in its detection and switching. Recent advancements showed the robustness of in-plane spontaneous polarization and related domain structure down to the monolayer (2D limit) at room temperatures contrary to that expected for out-of-plane polarization [15,16]. Thus, in-plane polarized domain configuration offers an opportunity to overcome the main hurdle of depolarization fields, reduced $T_C$ in the low dimensions and hence an advancement over the presently available ferroelectric devices. In-plane polarized domains and DWs are hard to detect and manipulate. Finding a parameter for non-contact control of the in-plane polarized domain configuration and DW movement in ferroelectrics is important for functional devices. The optical control and detection of DW movement are technologically promising for overcoming the conventional destructive readout methods and hence suitable for optoelectronic applications employing DWs [5,17,18,19,20]. In particular, the in-plane polarized domain configuration is attractive for devices, where domain configuration can be manipulated optically and remain unaffected by the out-of-plane depolarization fields [11,14,21,22,23]. A non-destructive control of in-plane polarized domains without any physical contact is still an open question and yet to be explored.

It is well known that the classical ferroelectric $BaTiO_3$ (BTO) consists of two types of domains, in-plane polarised *a*-domains and out-of-plane polarized *c*-domains [20]. Fig. 1(a) shows a schematic of the top view of *a*- and *c*- domains [24]. Apart from that, there is a possibility of formation of subdomains with 90° in-plane polarization alignment inside the *a*- domain, which can be defined as $a_1$- and $a_2$- domains (fig. 1(b)) [14,24]. Presence of such in-plane polarized 90º domains were previously suggested in single-crystals of BTO [25,26], however, remained unexplored. Unlike *a*- and *c*-domains, in-plane polarized sub-domain structures cannot be identified from the topographic images.

In the present communication, we utilized the principles of the angle-resolved polarized Raman spectroscopy (ARPRS) [27,28,29,30] to unveil hidden in-plane $a_1/a_2$ sub-domains in the single crystals of BTO which is further verified using X-ray diffraction studies. Raman mapping studies directly visualized the reversible optical switching of the $a_1/a_2$ subdomains configuration and subsequent DWs movement due to light polarization detected as a function of the power of incident laser with below bandgap energy. Interestingly, similar optical control of DW movement is also shown in the non-polar ferroelastic LaAlO$_3$ (LAO) crystal which implies that long-range ferroelectricity is not an essential ingredient for the observed DWs movement. Using experiments and first-principle calculations, we illustrate the crucial role of flexoelectricity in photoexcitation and subsequent electronic screening at the DWs resulting in the optical control of domain configuration and DW movement. The natural flexoelectricity-assisted optical domain switching and DW movement in ferroelectric and ferroelastic materials is observed for the first time. The observations put forward ferroelastic materials as a potential candidate for next generation DW nanoelectronics.

**Results**

   **1. Polarized Raman mapping of in-plane domain structure**

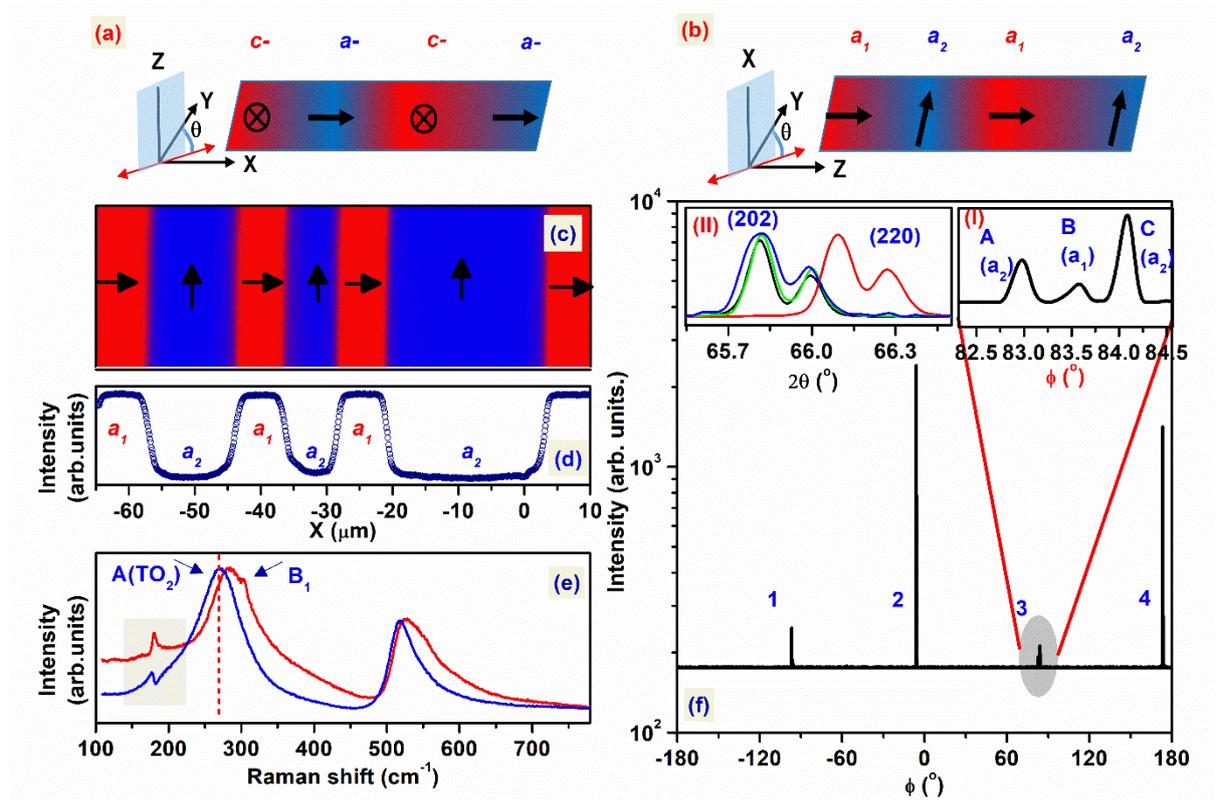

Figure 1 Identification of in-plane domain structure of single-crystal BTO: (a) Schematic of *a-c* domain structure (top view) in BTO; (b) Schematic of the in-plane polarized domain structure consisting of $a_1$ and $a_2$ subdomains; arrows and crosses mark the direction of the polarization. (c) Raman map constructed using a spatial variation of the intensity ratio of B$_1$ and A(TO$_2$) mode ($I^{\parallel}_{B_1}/I^{\parallel}_{A(TO_2)}$), collected in parallel polarization configuration on an *a*-domain. (d) Variation of the $I^{\parallel}_{B_1}/I^{\parallel}_{A(TO_2)}$ across the domain structure. (e) Polarized Raman spectrum of BTO corresponding to the $a_1$ (red) and $a_2$ (blue) subdomains. (f) Off-specular X-ray diffraction study of BTO crystal. Inset (I) shows a magnified view of feature 3. Inset (II) shows an asymmetric θ-2θ scan for various features in 2,3.

The domain structure and orientation of polarization in single-crystal BTO was identified using ARPRS studies. The ARPRS involve the examination of changes in the intensity of a Raman mode as a function of in-plane rotation. For parallel polarization geometry, the Raman intensity equation (See section 1, fig. S1 and related discussions in supplementary material[31]) suggests that the intensity of $B_1$ mode should change according to $I^{\parallel}_{B_1} \propto [c\sin^2\theta]^2$ as a function of in-plane rotation angle ($\theta$) for a single $a$-domain. Here, $\theta$ is the angle between the incident light polarization (**E**) and in-plane electric polarization (**P**) in the BTO crystal. For **E**∥**P** the angle $\theta=0°$ which gives $I^{\parallel}_{B_1} = 0$ while for **E**⊥**P**, the angle $\theta=90°$ which leads to $I^{\parallel}_{B_1} \propto c^2$. For the ARPRS study, we identified a single $a$-domain by the absence of E(LO$_4$)/A(LO$_3$) mode around 720 cm$^{-1}$ [20,28,32]. Then, we carried out ARPRS studies on this single $a$-domain. The experimental results were found be to in good agreement with the Raman intensity equation given above (fig. S1(a,b)). These results indicate that the absence of $B_1$ mode in the polarized Raman spectrum can directly reveal the direction of the **P** inside an $a$-domain.

We use this information to align our BTO crystal in such a way that **E**∥**P** and then a polarized Raman map was acquired. Raman map is constructed using the spatial variation of the intensity ratio of $B_1$ and A(TO$_2$) mode i.e. $I^{\parallel}_{B_1}/I^{\parallel}_{A(TO_2)}$. A representative Raman map over a 75×1 μm$^2$ region is shown in fig. 1(c). Fig. 1(d) shows a line scan of $I^{\parallel}_{B_1}/I^{\parallel}_{A(TO_2)}$ across the scanned area. It is seen that $I^{\parallel}_{B_1}/I^{\parallel}_{A(TO_2)}$ is varying alternatively over the scanned line. The polarized Raman spectra corresponding to the minima and maxima of $I^{\parallel}_{B_1}/I^{\parallel}_{A(TO_2)}$ (corresponding to blue and red regions in polarized Raman map Fig. 1(c), respectively) are shown in Fig. 1(e). These Raman spectra are identical to the Raman spectra obtained in the present ARPRS study for **E**∥**P** and **E**⊥**P**, (shown in fig. S1(a)), respectively. Furthermore, the angular-dependence of $I^{\parallel}_{B_1}/I^{\parallel}_{A(TO_2)}$ on these two regions showed a 90° phase difference (See supplementary fig. S2 [31]), which implies an alternative 90° alignment of **P**. It is also noted that the 720 cm$^{-1}$ mode is absent in both cases, which confirms the overall $a$-domain structure in the whole scanned area. These alternative regions inside the $a$-domain are referred as $a_1$ (**E**⊥**P**) and $a_2$ (**E**∥**P**) subdomains for ease, in which **P** lies within the plane but is oriented at 90° relative to each other [14,24].

2. **X-ray diffraction**

The presence of the in-plane $a_1/a_2$ subdomain configuration has been further verified using off specular X-ray diffraction study. Fig 1(f) shows the φ-scan on (202) reflection. The sample was tilted at an angle (χ) of 44.8° relative to the [100] direction. For a purely $a$-domain in tetragonal BTO, (100) is a two-fold axis, therefore two reflections are expected in the whole 360° φ-scan for (202) reflection. However, in our φ-scan study it is seen that apart from the two highly intense peaks (feature 2 and 4) at -6° and 174°, respectively, two additional set of peaks (feature 1 and 3) are appearing at -96° and 84°, respectively. Inset I of fig. 1(f) shows the magnified view of feature 3 which consists of three peaks (A, B, C). The peak B is at exactly 90° to features 2 and 4. While A and C peaks are separated by ~0.55° apart from the peak B. To confirm the origin of feature 2 and peaks (A,B,C) in feature 3, we have performed the θ-2θ scan at φ values corresponding to these peaks. The inset II of fig. 1(f) shows the normalized θ-2θ scan carried out at φ value of feature 2 (in blue) and feature 3 (A in black, B in red and C in green), respectively. We found that feature B corresponds to the (220) reflection and the A, C peaks of feature 3 and feature 2 originated from the (202) reflection. Ideally, (220) reflection should not appear in the φ scan on (202) reflections, however, it is appearing because of the small difference between 2θ and χ values of (202) and (220) planes and which is under our experimental resolution limit. The observed fourfold orientation of the (202) plane is possible only when $a$-domain consists of two in-plane 90° domain structures ($a_1/a_2$ subdomains). Furthermore, the separation between the A and C peaks (~1.1°) of features 1 and 3 signifies that the subdomains are not oriented exactly at 90° but rather they are at 89.45° due to the tetragonality which further supports existence of the ~90° domain structures [33].

3. **Light-induced DW movement**

Now, we examine the influence of incident laser power on the observed in-plane polarized subdomain structure using polarized Raman mapping. During these experiments, the **E**∥**P** in $a_2$ domains (blue color) and **E**⊥**P** in $a_1$ domain (red color). Fig. 2(a-e) show the changes in the domain configuration ($a_2$-$a_1$ subdomain conversion) and associated DW movements as a function of the laser power (0.06-6 mW). Fig. 2(f) shows the movement of the DW at positions 1-4 in fig 2(a) as a function of laser power. It is seen that the DW movement is different for different domain, maximum shift observed at position 2 (~5μm).

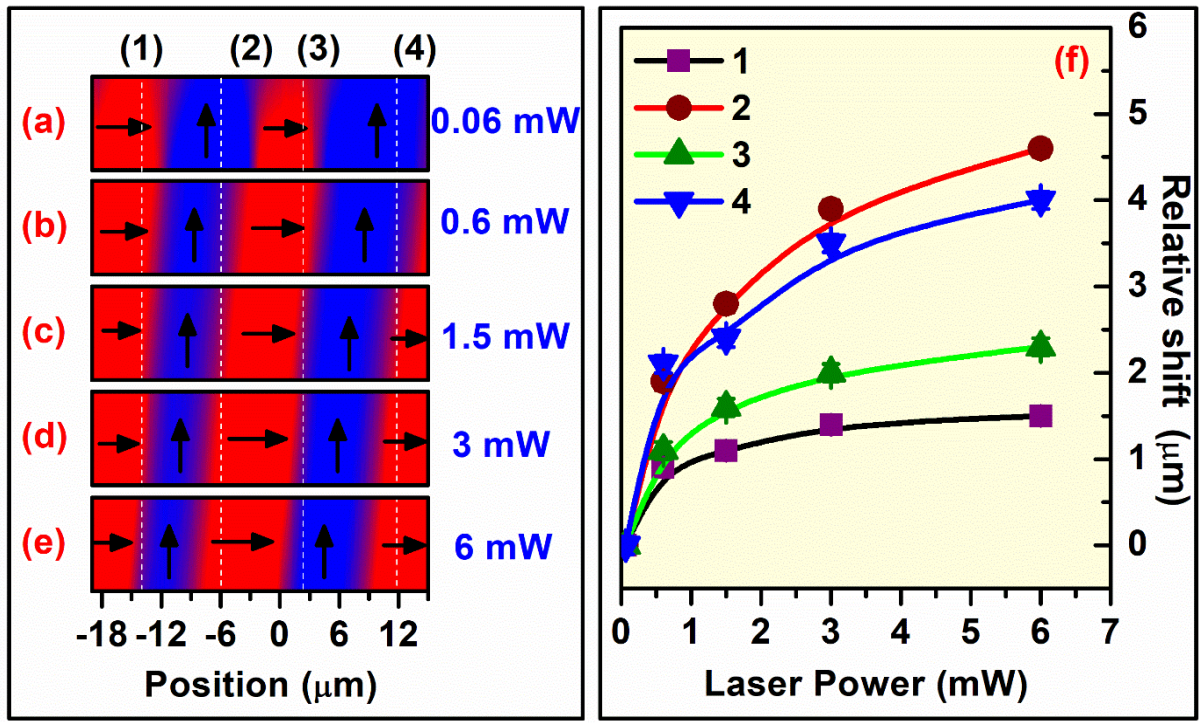

Figure 2 Domain reconfiguration and DW movement under polarized light: (a-e) Raman mapping of $a_1$-$a_2$ subdomain configuration under different incident laser power f) Relative movement of the DW extracted from Raman maps (a-e) at position 1-4 as a function of incident laser power.

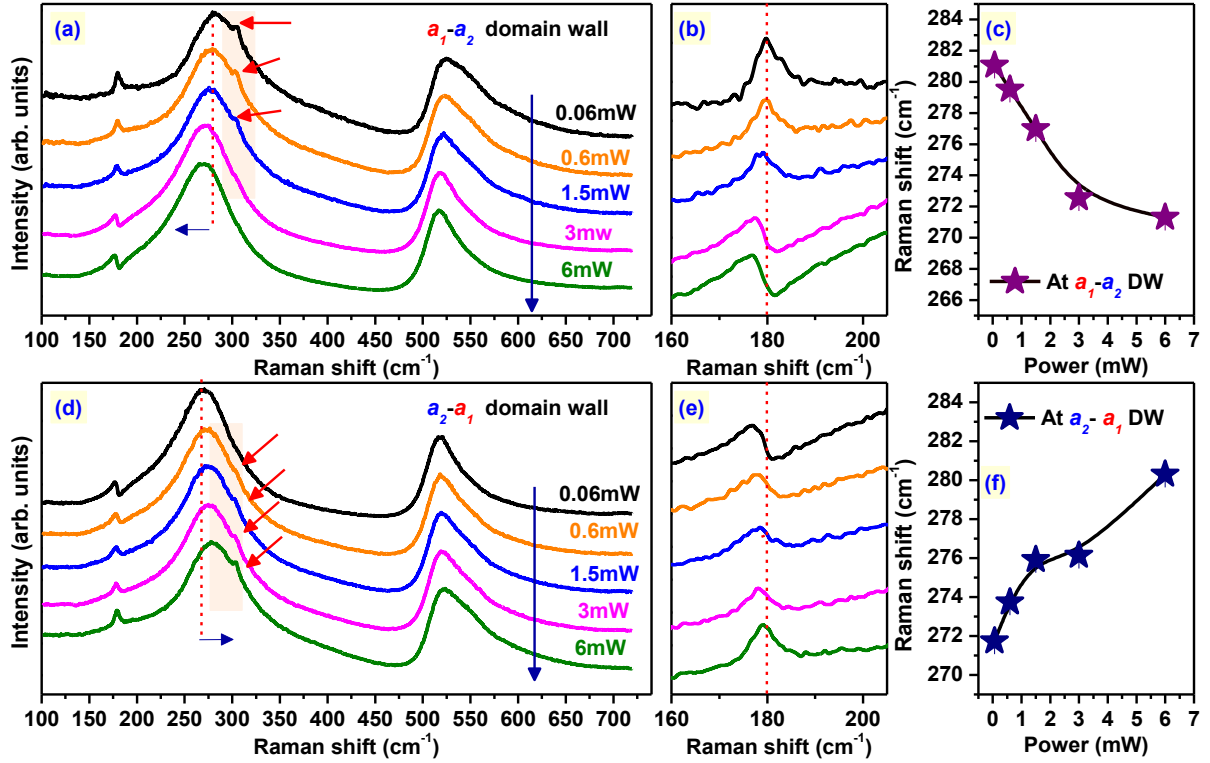

Figure 3 | Effect of the laser power on the Raman spectra of BTO: (a-c) Laser power dependent variation in Raman spectrum measured at $a_1$-$a_2$ DW (position 1 in fig. 2) and (d-f) for $a_2$-$a_1$ DW (at position 2 in fig. 2).The comparison of the laser power-dependent Raman shift at the $a_1$-$a_2$ and $a_2$-$a_1$ DWs, is shown in (c,f), respectively.

To elucidate the local structural modifications, we examined the polarized Raman spectrum near $a_1$-$a_2$ (position 1 in fig. 2) and $a_2$-$a_1$ (position 2 in fig. 2) DWs as a function of incident laser power. Fig 3(a) shows the Raman spectrum as a function of laser power near $a_1$-$a_2$ DW. We note here that the line shape of the A(TO$_1$) Raman mode ~180 cm$^{-1}$ was found to switch from symmetric Lorentzian to asymmetric Fano line (fig. 3(b)) and the A(TO$_2$) mode showed a remarkable softening (281 to 272 cm$^{-1}$ in fig. 3(c)) on increasing the power. The changes in the A(TO$_2$) mode position is indicative of changes in the Ti-O vibrational frequency reflecting modification in local polarization [20,27,34]. The change observed in the line shape of A(TO$_1$) mode around 180 cm$^{-1}$ also indicates towards modification in the local polarization [34]. Completely reverse effect was seen near the $a_2$-$a_1$ DW, the Raman mode was found to harden (272 to 280 cm$^{-1}$) fig. 3(d-f) and the asymmetric Fano line convert to the Lorentzian shape on increasing laser power (fig. 3(d,e) ). Also, the intensity of the B$_1$ mode (marked by red arrow in fig. 3(a,d)) near the DW regions was observed to change which indicates a local polarization rotation resulting in the growth of $a_1$ subdomains at the expense of $a_2$ subdomains along with a shift in DW. Interestingly, the Raman spectrum inside an individual subdomain was found to be invariant due to incident laser power variation (fig. S3,S4 [31]) indicating the absence of change in the local temperature, lattice strain, and polarization inside the subdomains.

4. **Reversible switching of domain structure**

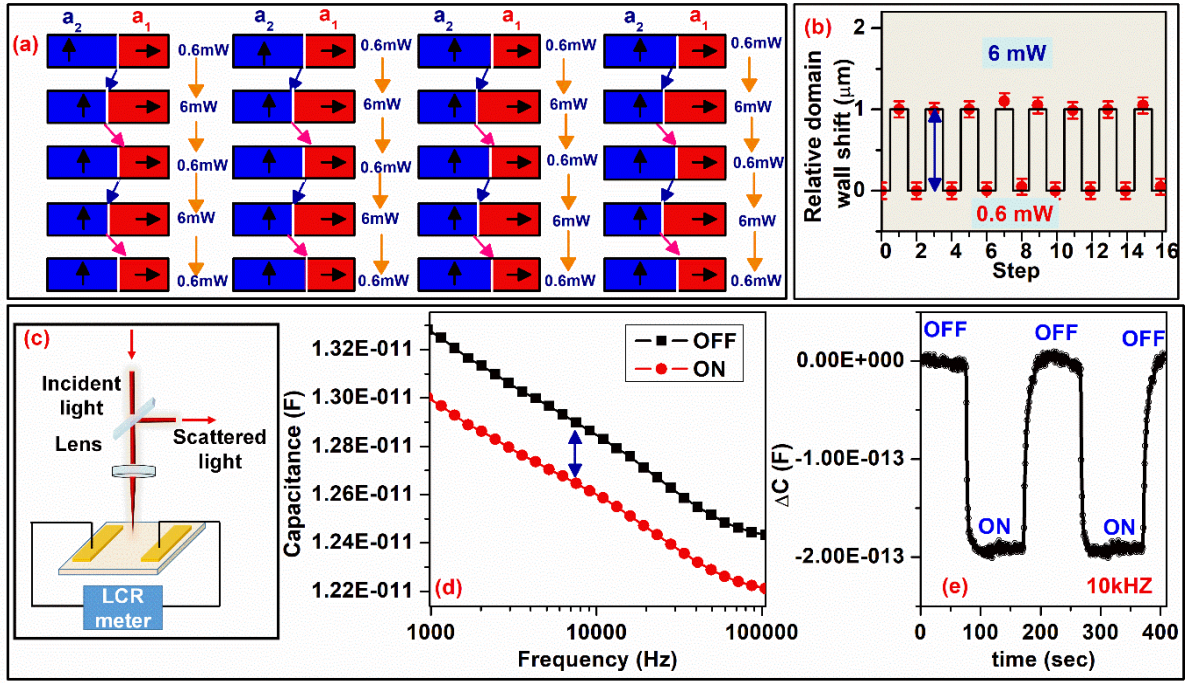

Figure 4 Reversible optical control of domain configuration and its effect on macroscopic dielectric response: (a) Optical switching of $a_1/a_2$ subdomains and DW movement with laser power swapping between 0.6 and 6 mW. (b) The relative shift of DW quantified from Raman maps shown in fig. 4(a). (c) Schematic of the experimental geometry of laser dependent dielectric studies. (d) Frequency-dependent and (e) Time-dependent variation of in-plane dielectric constant under laser light illumination (6 mW).

Practical applications demand reversibility of the phenomena. To test the reversibility of the observed DW movement, we performed a switching experiment by swapping the laser intensity between 0.6 and 6 mW and recorded the Raman map under different incident laser power. The experiment was repeated for several cycles and the observed changes for these cycles are shown in fig. 4(a). The white line marks the DW. It is noticeable in fig. 4(a) that on switching the laser power from 0.6 to 6 mW, the DW shifts to the left along with the partial conversion of $a_2$ subdomain into the $a_1$ subdomain. On switching back the laser power, the DW shifts to the right and regains the initial domain configuration. Therefore, it is seen that the DW position is reversibly switched (1±0.1 μm) with laser power as shown in fig. 4(b). Further, to see the macroscopic effect of $a_1/a_2$ subdomains switching, we performed in-plane dielectric measurement under light illumination. Fig. 4(c,d,e) shows the experimental geometry, frequency and time-dependent dielectric response under light ON-OFF, respectively. The reversible changes observed in Raman map studies and in the in-plane dielectric measurements proves the optical control of domain structure and subsequent changes in macroscopic response.

5. **Light polarization-dependent DW movement**

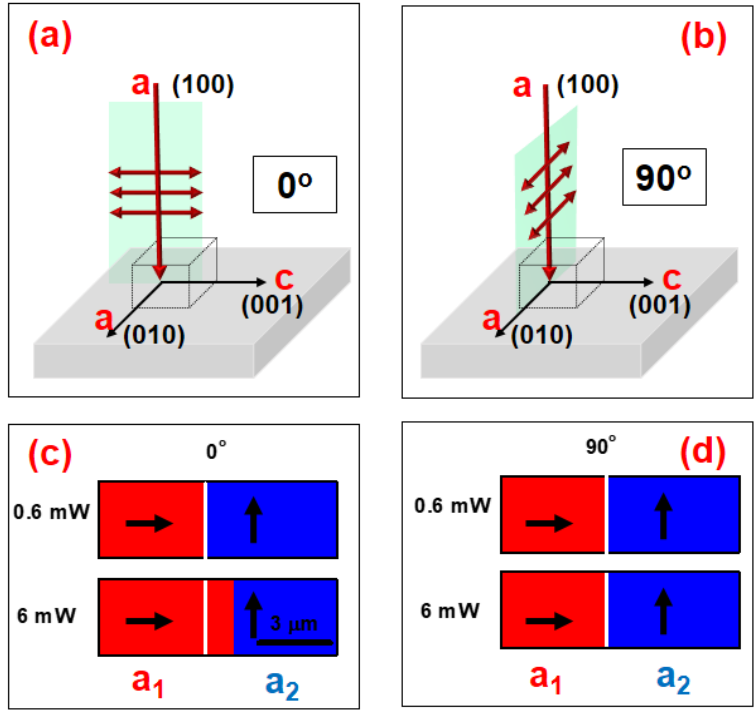

figure 5 Light polarization dependent DW movement: Experimental geometry of the light polarization-dependent Raman mapping for (a) 0° (b) 90° light polarization and corresponding Raman maps for 0.6 mW and 6 mW laser power using 633nm excitation are presented in (c) and (d).

To address the physical mechanism responsible for this effect, we first rule out the thermal contributions [20,35,36]. Observation of laser power independence of the Raman shift inside the subdomains and the opposite effects near the two DWs shown in fig. 3 (and S3, S4) negates the possibility of thermal dilation effects due to laser power, as thermal effects are supposed to create expansion in the lattice which should result in mode softening. For further confirmation, we examined the wavelength and laser polarization dependency. Fig. 5(a,b) shows the experimental geometry of light polarization-dependent studies with light polarization θ=0°, and 90°. The Raman maps collected for 0.6 mW and 6 mW laser power for 633 nm excitation in these two configurations are shown in fig. 5(c) for 0°, and in fig. 5(d) for 90° light polarization. It is noted that when light polarization is along the electric polarization (θ=0°), the $a_1$ sub-domain grows and shows a DW movement of ~1.4 μm with laser power while it is negligible for θ=90°. The results obtained using 473nm laser were found to be identical. These observations indicate selective control of the DW movement fundamentally by incident light polarization and tuned by the power of the incident laser light. The angular dependency clearly rules out thermal dilation and suggests electronic screening (photovoltaic) as the driving force as observed in our recent work [20]. However, the following questions need further attention: 1) why the domain structure is not much affected with laser power for θ=90°? is it related to the role of long-range ferroelectricity? 2) What is the origin of the process of below bandgap photoexcitaion ($E_{Laser}$~1.92eV (632.8 nm), 2.67eV (473 nm)) and its role in DW movement?

6. DW movement in ferroelastic LaAlO$_3$

In order to address these issues, we performed similar experiments on a non-polar, but ferroelastic LAO ($E_g$>5 eV) single crystal having an elastic domain structure [37]. Fig. 6(a-d) shows a Raman map constructed using intensity of $A_g$ mode of the LAO measured under 0.6 to 6 mW incident laser power (see supplementary materials fig. S5, S6 for details of Raman mapping [31]). A relative change in the domain size due to incident laser power is reflected in terms of an increased red region. Fig. 6(e) shows line scans across the domains under 0.6 mW (black) and 6 mW (red) of incident laser power, respectively, depicting the growth of a particular domain at the expense of the other on increasing laser power. The non-thermal nature of this event is established by performing light polarization dependent experiments on LAO (fig. 6(f,g)). Further, the Raman mode position inside individual domains was found to remain unchanged as a function of laser power similar to BTO indicating the absence of change in the local temperature, and lattice strain inside the domains. It may further be noted that the relative shift of DW is different at different walls as has been observed in BTO which can be attributed to different domain sizes.

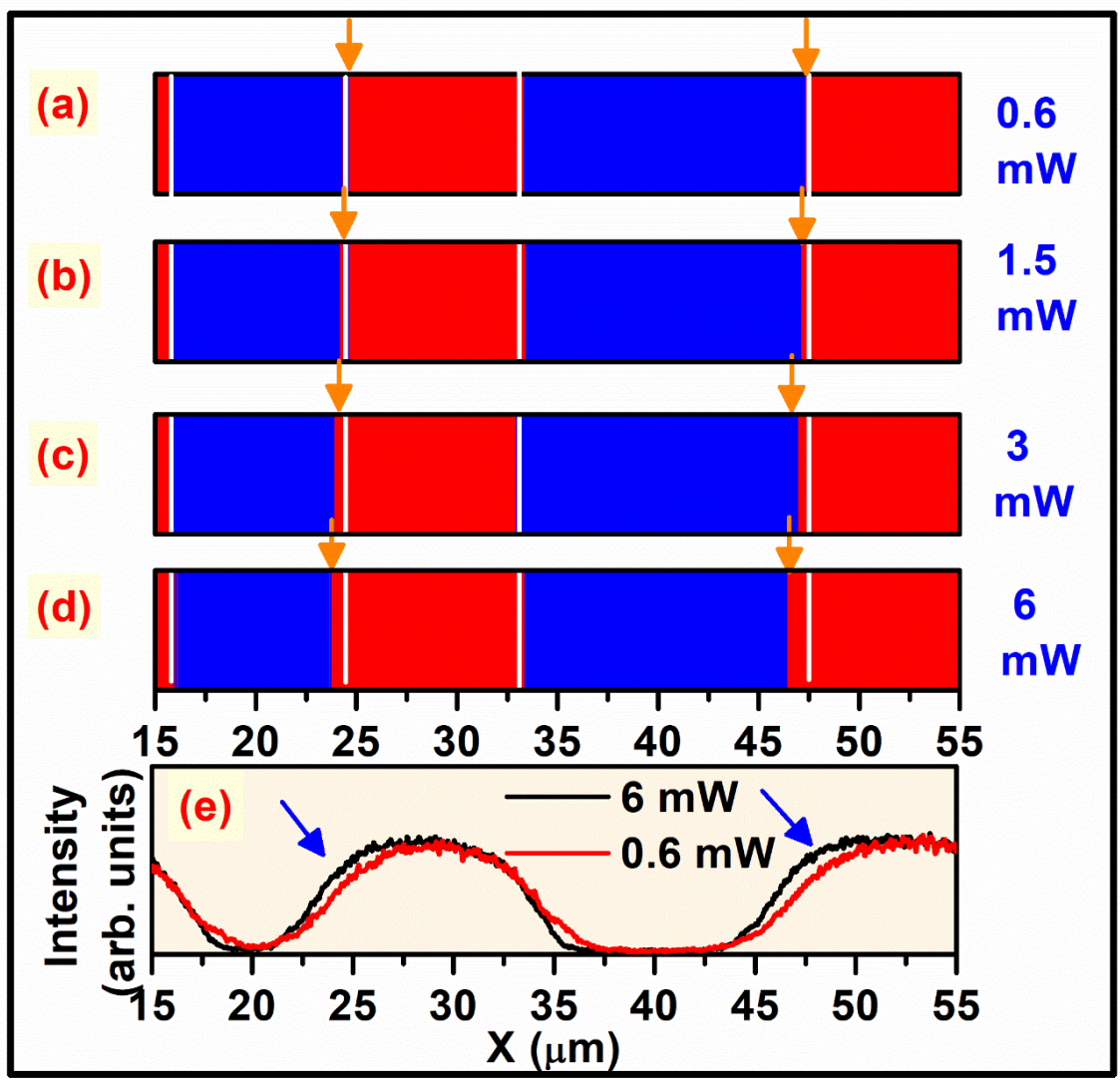

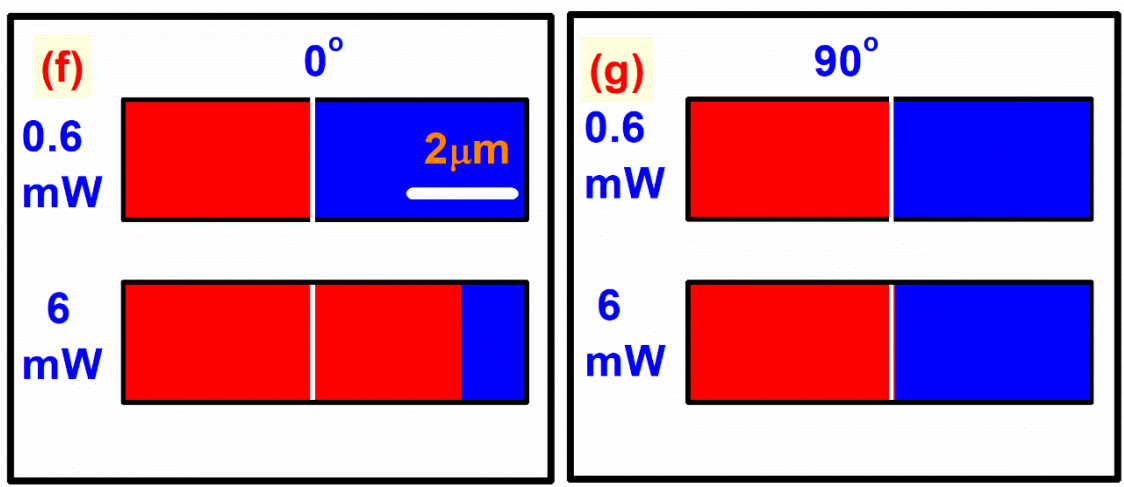

Figure 6 DW movement in LAO under polarized light: (a-d) DW motion in LAO in terms of spatial variation of $A_g$ mode intensity (~126 cm$^{-1}$) as a function of incident laser power variation from 0.6 to 6 mW. Arrows mark the position of the DW, white line marks the position of DW for 0.6 mW laser intensity. e) Line scan of Raman intensity across domains for incident laser power of 0.6 mW (black) and 6 mW (Red). f-g) Laser power-dependent DW movement under 0° and 90° light polarization configuration.

## 7. First principle calculations

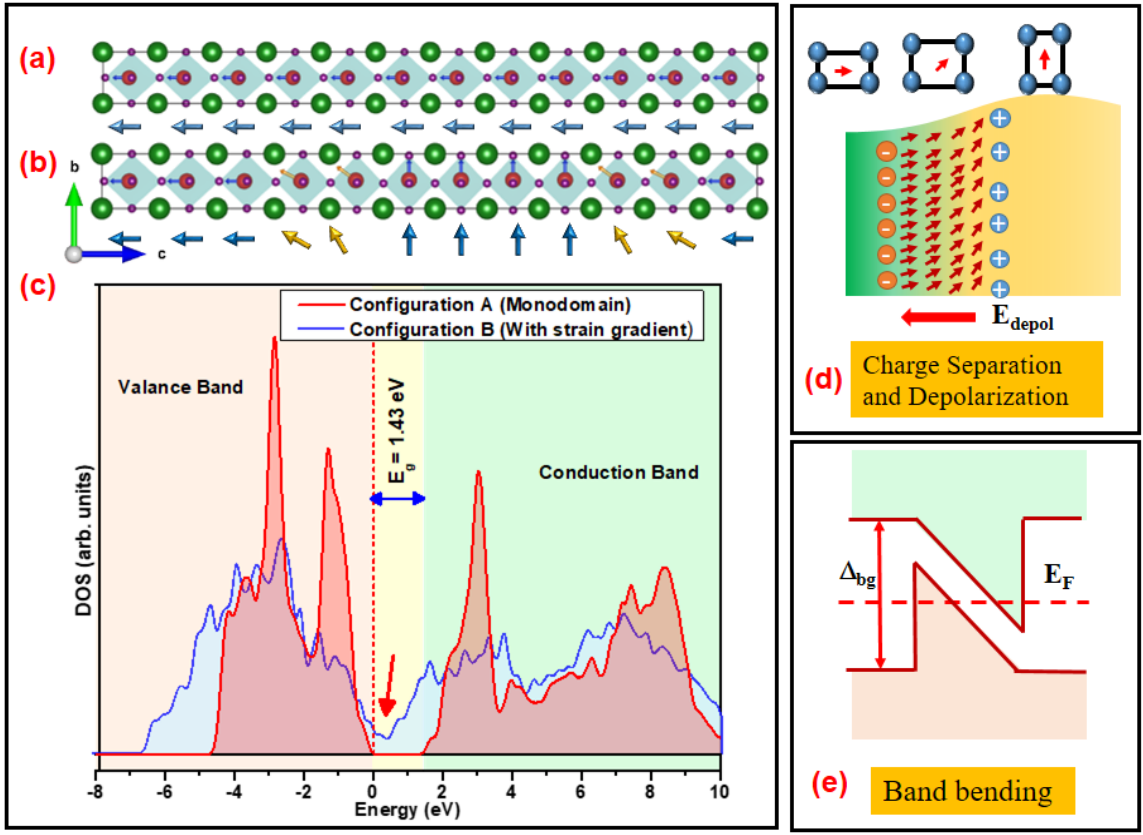

Figure 7 Electronic reconstruction due to strain gradient: Supercell configuration for the first principle calculation (a) monodomain (configuration A) (b) with strain gradient (configuration B), (constructed using VESTA software [38]). Arrow indicates the direction of polarization (c) Calculated density of state corresponding to supercell configuration A, and B shown in red and blue curves, respectively (the valence band maxima is shifted to 0 eV). Vertical dashed red line represent the Fermi level. Schematics representation of the Charge separation (d) and electronic reconstruction and band bending (e) due to the strain gradient.

The light polarization dependence of the DW movement with laser power in BTO indicates a correlation with the asymmetric potential and long-range feroelectricity in the system. However, observation of similar DW movement in non-polar LAO suggests that long-range ferroelectricity is not essential. The change in Raman spectra with laser power only near DWs indicate that the observed effect is linked with the DW itself. Despite the lack of long-range ferroelectricity, the strain-gradient induced polarization components i.e. flexoelectricity is a universal phenomenon of the structural DWs [39,40,41,42]. Ahluwalia et al [43] pointed out the role of flexoelectricity in the stabilization of the domain configuration. Furthermore, the flexoelectric coupling-induced polarization is known to alter the band structure [44] which leads to the reduced bandgap. The present Raman mapping studies clearly identified regions of strain gradient in the vicinity of the DWs in the BTO and LAO [See fig. S6,S7]. To understand the role of this strain-gradient on the electronic structure across the domain walls, we investigated the electronic structure in the tetragonal phase of BTO using first-principle calculations. We started with structural optimization of the BTO unit cell. The lattice parameter of the relaxed tetragonal-BTO unit cell was found to be a=4 Å and c=4.2 Å [32,34]. Using this optimized unit cell, we constructed the two configurations using a $1a \times 1a \times 12c$ supercell. In configuration A (fig. 7(a)), represents a monodomain with polarization along the *c*-axis of the supercell. Whereas, in configuration B (fig. 7(b)), the dipole moment in each unit cell is modulated to mimic an interface of two ferroelectric domains with polarization pointing towards the *c*- and *b*- axis. Each domain contains 4 unit cells. The strain-gradient can be applied in any arbitrary direction; however, that requires a much larger supercell with more complex geometries. Hence to make it simple and to investigate the effect of the strain-gradient on electronic structure qualitatively, we have implemented the strain gradient and domain interface by introducing a spatial gradient in the *c*-parameter (with a fixed *a*-parameter). We evaluated the electronic structure of these configurations using the first principle calculations (details of the supercells and calculations are given in supplementary materials [31]). The valence band maxima is set to 0 eV in all our calculations. The calculated bandgap in configuration A (i.e. Bulk tetragonal BTO) was found to be ~1.43 eV, while for configuration B, there is a finite density of state at Fermi level. To further see the effect of homogenous lattice strain on the bandgap, we performed additional calculations with a smaller *c*-parameter for configuration A and B. This decrement in the spontaneous strain was found to increase the bandgap to 1.57 eV for configuration A (fig. S9), while the configuration B showed similar finite DOS at the Fermi level. This clarifies that the strain gradient is responsible for the finite DOS at Fermi level. Next, we examined the layer resolved DOS (LDOS) of the configuration B. The layer resolved DOS revealed a clear relative shift in the energy bands across the regions of strain gradient (fig. S10).

The observations are in good agreement with the previous theoretical reports suggesting strain gradient induced local polarization and subsequent electronic reconstruction [44,45]. The bandgap reduction is experimentally observed even in uncharged DWs of $BiFeO_3$ and $Pb(Zr,Ti)O_3$ systems [9,46]. Fig. 7(d,e) show the schematic of the strain-gradient induced polarization and subsequent band alignment across DW region. Because of this electronic reconstruction, photoexcitation can occur for excitation energy below the bandgap. Finally, we emphasize that electronic reconstruction depends on the strength of the strain gradient [47], and may result in similar below bandgap photoinduced structural modulation previously observed in ref. [48,49] and in present experimental studies.

**Discussion**

The strain-gradient induces charge separation or local polarization ($P = \varepsilon. f_{eff} \frac{\partial u}{\partial z}$ [44]) which gives rise to a built-in local electric field given by $E_s = \frac{e}{4\pi\varepsilon_0 a}\frac{\partial u}{\partial z}$ [50] i.e. an asymmetric potential across the strain-gradient regions, mostly in DWs [51] (fig. 7). The charge separation or local polarization is a consequence of strain-gradient irrespective of long-range ferroelectricity. The electronic re-construction due to the presence of strain gradient (flexoelectric field) at DWs allows the photoexcitation process even from the below bandgap excitation. The strain-gradient induced local asymmetric field ($E_s$) effectively separates out the photo-excited carriers and hence, has angular dependence as a function of incident light polarization [52,53,54,55,56]. The photo-excited carriers screen the local polarization components, and renormalize the local electric field or strain-gradient [57]. Borisevich *et al* suggested emergence of DW instability from the renormalization of the strain-gradient [58]. Ahn *et al* had shown optically induced DW tilting due to screening from the photoexcited carriers [59]. Though the number of photoexcited carriers is small, however, because of the nanoscopic width of the DWs, the carrier density becomes significantly high. This creates very high local strain fields resulting in photo-induced lattice deformations (as observed in fig. 2,3) and subsequent domain reconfiguration [19]. Catalan *et al* had shown that strain-gradient or flexoelectric coupling is responsible for local polarization reorientation at 90° DWs in $PbTiO_3$ [33]. Hence, the flexoelectricity-assisted photostriction is the dominant cause for domain interconversion and associated DWs movement as a function of laser power (and polarization) observed in the present study. Our present study also gives experimental evidence of theoretically predicted flexoelectricity-driven DW movement in ferroelastic material without alteration of order parameters inside the domains [60].

**Conclusion**

This study establishes in-plane subdomain structure in BTO single crystal through Raman spectroscopy and X-ray diffraction studies. It is shown that the in-plane domains typically called *a*-domains in BTO may have 90° sub-domain structure ($a_1/a_2$). A non-thermal, reversible optical control of the in-plane domain configuration and DW movement is demonstrated in polar BTO and non-polar ferroelastic LAO single crystals. The detailed Raman mapping investigations provided a direct visualization of the optical tunability of the domain configuration and DW movement irrespective of absence or presence of long range ferroelectricity. The strain-gradient induced local polarization and subsequent electronic reconstruction was found to play an important role in the optical control of DW movement in non-ferroelectric systems. The microscopic mechanism of the optical control of DW movement is illustrated as photo-induced changes through flexoelectricity or "Flexo-photostriction". The observed unconventional flexoelectricity-driven DW movement in economical and readily available ferroelastic material provide a new understanding for DW nanoelectronic devices with unprecedented opportunities without need of long range ferroelectricity.